\documentclass[12pt]{article}
\usepackage{amsfonts,epsfig,latexsym,amscd}
\usepackage{amssymb}

\newcommand{\R}{{\mathbb{R}}}
\newcommand{\Z}{{\mathbb{Z}}}

\newcommand{\I}{{\mathbb{I}}}

\newcommand{\be}{\begin{equation}}
\newcommand{\ee}{\end{equation}}
\newcommand{\bea}{\begin{eqnarray}}
\newcommand{\eea}{\end{eqnarray}}
\newcommand{\bean}{\begin{eqnarray*}}
\newcommand{\rt}{\tilde r}

\newcommand{\eean}{\end{eqnarray*}}
\font\upright=cmu10 scaled\magstep1

\newcommand{\PP}{\hbox{\upright\rlap{I}\kern 1.5pt P}}

\newcommand{\identity}{{\upright\rlap{1}\kern 2.0pt 1}}

\newcommand{\HH}{\mbox{\hbox{\upright\rlap{I}\kern 1.7pt H}}}

\newcommand{\fr}{\frac}
\newcommand{\lm}{\lambda}
\newcommand{\ra}{\rightarrow}

\newcommand{\sg}{\sigma}

\newcommand{\pr}{\partial}
\newcommand{\x}{ {\bf x} }
\newcommand{\hs}{\hspace{5mm}}

\newcommand{\dg}{\dagger}

\newcommand{\ve}{\varepsilon}
\newcommand{\acc}{\\[3mm]}

\newcommand{\vv}{{\bf v}}

\begin{document}
\setcounter{page}{0}
\begin{titlepage}
\strut\hfill
\vspace{0mm}
\begin{center}

{\large\bf  Weyl Equation and (Non)-Commutative $SU(n+1)$ BPS Monopoles}
\vspace{12mm}

{\bf Anastasia Doikou${}^*$ \ and \ Theodora Ioannidou${}^\dg$}
\\[8mm]
\noindent ${}^*${\footnotesize Department of Engineering Sciences, University of Patras,
GR-26500 Patras, Greece }\\
{\footnotesize E-mail: adoikou@upatras.gr}
\\[8mm]
\noindent ${}^\dg${\footnotesize Department of Mathematics, Physics and Computational Sciences, Faculty of Engineering,\\
Aristotle University of Thessaloniki, GR-54124 Thessaloniki, Greece }\\
{\footnotesize E-mail: ti3@auth.gr}

\vspace{12mm}

\begin{abstract}

We apply the ADHMN construction to obtain the $SU(n+1)$ (for generic values of $n$) spherically
symmetric BPS monopoles with minimal symmetry breaking. In particular, the
problem simplifies by solving the Weyl equation, leading to a set
of coupled equations, whose
solutions are expressed in terms of the Whittaker functions.
Next, this construction is generalized for {\it non-commutative}
$SU(n+1)$ BPS monopoles, where  the corresponding solutions
are given in terms of the Heun B functions.

\noindent

\end{abstract}

\end{center}
\end{titlepage}

\tableofcontents
\section{Introduction}

Bogomolny-Prasad-Sommerfield (BPS) monopoles are topological solitons in a Yang-Mills-Higgs
gauge theory in three space dimensions.
The equation for static monopoles is integrable, and thus a variety of techniques are available for constructing and 
studying the monopole solutions.
Direct construction of solutions of the Bogomolny equation with monopole number greater than one is a difficult task
with the exception of spherical symmetry, (see, for example, Ref \cite{BW}-\cite{IS} and References therein).
To bypass this difficulty a number of intriguing ideas have been put forward rendering this complicated problem more tractable.

A powerful approach introduced by Nahm \cite{Nahm} is the so-called Atiyah-Drinfeld-Hitchin-Manin-Nahm (ADHMN) construction.
The Atiyah-Drinfeld-Hitchin-Manin (ADHM) construction \cite{ADHM}, allows the construction of instantons in terms of
linear algebras in a vector space, whose dimension is related to the instanton number. Since monopoles correspond to infinite
action instantons, then an adaptation of the ADHM construction involving an infinite dimensional vector space might be also possible.
Nahm was able to formulate such an adaption in the ADHMN construction.
To perform this construction a nonlinear ordinary differential equation (i.e. the Nahm equation), must be solved and its solutions 
(i.e. the Nahm data), used to  define the Weyl equation.

The Nahm equations provide a system of non-linear ordinary differential equations
\be \fr{dT_i(s)}{ds}=\fr{1}{2}\, \ve_{ijk}\,[T_j(s),\ T_k(s)]\label{Nahm}\ee
for three $n\times n$ anti-hermitian matrices $T_i(s)$ (the so-called Nahm data) of functions
of the variable $s$, where $n$ is the magnetic charge of the BPS monopole configuration.
The tensor $\ve_{ijk}$ is the totally antisymmetric tensor.
These equations may be obtained from the self-dual Yang-Mills equations in four dimensions by imposing
translation invariance in three dimensions.
Thus, equations (\ref{Nahm}) are completely integrable, and can be solved in terms of abelian functions.

In the ADHMN approach, the construction of  $SU(n+1)$ monopole solutions to the Bogomolny equation with
topological charge $n$ is translated to the following problem which is known as
the inverse Nahm transform \cite{Nahm}.

Finding the Nahm data effectively solves the nonlinear part of the monopole construction, however in order to
calculate the fields themselves the linear part of the ADHMN construction must be implemented. Given the Nahm data for
a $n$-monopole the one-dimensional Weyl equation
\be
\left( \I_{2n}\fr{d}{ds}-\I_n\otimes x_j \sg_j +iT_j\otimes\sg_j\right)\vv({\bf x},s)=0 \label{Weyl}\ee
for the complex $2n$-vector $\vv(\x,s)$, must be solved. $\I_n$ denotes the $n\times n$ identity matrix
and ${\bf x}=(x_1,x_2,x_3)$ is the position in space at which the monopole fields are to be calculated.
Let us choose an orthonormal basis for these solutions, satisfying
\be
\int \hat \vv^\dg \hat \vv \,ds=\I. \label{wnor}\ee
Given $\hat \vv(\x,s)$, the normalized vector computed from (\ref{Weyl}) and (\ref{wnor}), the Higgs field $\Phi$ and gauge 
potential $A_i$ are given by
\bea
\Phi&=&-i\int s\, \hat \vv^\dg \hat \vv \,ds,\label{Higgs}\\
A_i&=&\int \hat \vv^\dg \,\pr_i \hat \vv\, ds,
\eea
where the integrations are to be performed over the range spanned by the minimum and maximum
eigenvalues of the asymptotic form of $\Phi$. Then the corresponding gauge and Higgs fields satisfy
the Bogomolny equation and the boundary conditions for a charge $n$ monopole, and they are smooth
functions of $\x$. Although many general results have been obtained few explicit solutions are known.
Specifically, many solutions of the Bogomolny equation have been discovered using the inverse Nahm transform,
however this does not mean that analytic expressions for the gauge and Higgs fields are known.

One of the main aims of the present investigation is the construction of charge $n$, $SU(n+1)$ monopoles with symmetry 
breaking from $U(n)$.
This means that the asymptotic value of the Higgs field has $n$ equal eigenvalues; which is also referred to as minimal 
symmetry breaking.
Moreover, we construct the non-commutative version of our solutions.
As was recently shown, quantum field theories in non-commutative space-time naturally arise as a decoupling limit of
the world volume dynamics of D-branes in a constant Neveu-Schwarz{\Large -}Neveu-Schwarz (NS-NS) two-form background \cite{SW}.
In particular, we explicitly construct the solutions of the non-commutative BPS equations to the linear order of the
non-commutativity scale using a deformed (non-commutative) version of the Nahm equation.

\section{Weyl Equation and $SU(n+1)$ BPS Monopoles}

In this section, the  solutions of the Weyl equation for spherically symmetric $SU(n+1)$ BPS monopoles (for generic values
of $n$)
in the minim symmetry breaking case  are obtained explicitly. By spherical symmetry we simply mean here, that one has to 
deal only with the $r$ dependence,
i.e. no azimuthal dependence is considered.
To our knowledge the only work within this spirit is
presented in \cite{dancer,HS}, but again only for particular values of $n$ (i.e. $n=2$ and $n=3$, respectively).

In the minimal symmetry breaking case the Nahm data $T_i$ are defined on
a single interval $[-n,1]$ with the only pole occurring at the left-hand end of the interval. This allows a construction
 of Nahm data for charge $n$ monopoles in a minimally broken $SU(n+1)$ theory in terms of rescaled Nahm data for $SU(2)$ 
 monopoles, where the rescaling moves the second pole in Nahm data outside the interval. For convenience we shift $s$ so 
 that the Nahm data are defined in the interval, $[0,n+1]$.
Finally, the boundary conditions require that this representation in the unique irreducible $n$-dimensional representation
of $su(2)$ (see e.g. \cite{msbook} for a more detailed discussion).
Thus, the Nahm data
can be cast as
\be
T_i =-{i\over 2}\,f_i\,\tau_i, \hs i=1,\ 2,\ 3 \label{data}
\ee
where $\tau_i$'s form  the $n$-dimensional representation of $SU(2)$ and satisfy
\be
[\tau_i,\ \tau_j] = 2i \varepsilon_{ijk}\tau_k.
\ee
The $n$-dimensional representation of $SU(2)$ is of the form
\be
\tau_1 = \sum_{k=1}^{n-1} C_k \left(e^{(n)}_{k k+1}+ e^{(n)}_{k+1 k}\right),
\hs \tau_2=i\sum_{k=1}^{n-1} C_k \left( e^{(n)}_{k+1 k}- e^{(n)}_{k k+1}\right),\hs
\tau_3 = \sum_{k=1}^na_k\, e^{(n)}_{kk}
\ee
where $e^{(n)}_{ij}$ are $n\times n$ matrices defined by: $\left(e^{(n)}_{ij}\right)_{kl} = \delta_{ik} \,\delta_{jl}$ and
\be
a_k = n+1 -2k, \hs \hs C_k =\sqrt{k\,(n-k)}.
\ee
The Nahm data  for $SU(n)$ spherically symmetric monopole of charge $n-1$ are given by (\ref{data}) where $~f_i = f= -{1\over s}$
(see also e.g. \cite{dancer, msbook}). Since the monopole with this Nahm data is spherically symmetric,
we may choose $(x_1,x_2,x_3)=(0,0,r)$, and assume that the vector $\vv(\x,s)$ is of the form
\be
\vv(\x,s) = \sum_{l=1}^n h_l(r,s)\ \hat e^{(n)}_{l} \otimes \left(b_1(r,s)
\ \hat e_{1}^{(2)} + b_2(r,s)\ \hat e_{2}^{(2)}\right)
\ee
where $\hat e_{k}^{(n)}$ is the $n$-dimensional column vector with $1$ at the position $k\in \Z^+$ and $0$ elsewhere,
i.e. the standard basis of $\R^n$.

Then the Weyl equation (\ref{Weyl}) becomes
\bea
\Big [{d\over ds} +{f\over 2} \sum_{k=1}^{n-1}\, C_k\left(e_{k k+1}^{(n)} + e^{(n)}_{k+1 k}\right) \otimes \left(e_{12}^{(2)}
+ e_{21}^{(2)}\right) -
{f\over 2} \sum_{k=1}^{n-1}C_k \left(e^{(n)}_{k+1 k}-e^{(n)}_{kk+1}\right)\otimes\left(e_{21}^{(2)} -e_{12}^{(2)}\right)
&&\nonumber\\
+\,{f\over 2}\sum_{k=1}^n a_k\, e_{kk}^{(n)} \otimes \left(e_{11}^{(2)} - e_{22}^{(2)}\right) - r \,\I \otimes
\left(e_{11}^{(2)} -
e_{22}^{(2)}\right)\Big ]
\sum_{l=1}^n h_l\, \hat e_{l}^{(n)} \otimes \left(b_1 \hat e_{1}^{(2)} + b_2\ \hat e_{2}^{(2)}\right) =0. \hs  \
&&\label{w1}
\eea
To proceed with our computation we exploit the following properties:
\be
e^{(n)}_{ij}\ e^{(n)}_{kl} = \delta_{kj} e_{il}^{(n)},\hs \hs e_{ij}^{(n)}\ \hat e_k^{(n)} = \delta_{jk}\ \hat e^{(n)}_i.
\ee
With the use of the latter identities and after setting
\be
u_l(r,s) = h_l(r,s)\,b_1(r,s), \hs w_l(r,s) = h_l(r,s)\,b_2(r,s),
\ee
equation (\ref{w1}) is equivalent to the following first-order system of differential equations
\bea
&&\dot{u_1} -\left({1\over 2s}\,a_1+r\right)u_1 =0, \nonumber\\
&& \dot{u}_{k+1} -{1\over s} C_{k}\,w_{k} -\left({1\over 2s}\, a_{k+1} +r\right) u_{k+1}=0, \nonumber\\
&& \dot{w}_{k} -{1\over s}\, C_{k} \,u_{k+1} + \left({1\over 2s} \,a_{k} +r\right) w_{k} =0,\nonumber\\
&& \dot{w}_n + ({1\over 2s}\, a_n +r)w_n =0, \label{equations}
\eea
where  $k=1,2, \dots, n-1$.
Here, $\dot{u_i}$ and $\dot{w_i}$ for $i=1,\dots,n$ are the total derivatives of the functions $u_i(r,s)$ and $w_i(r,s)$
with respect to the argument $s$.
Solving these equations is the first step in reconstructing a solution of the Bogomolny equations from Nahm data.
Then the problem of recovering the Higgs and gauge fields is linear.

Let us now solve these equations. Assume first that $r \neq 0$.
The first and last equations may be immediately integrated and their solutions are equal to
\be
u_1 = \kappa_1(r) \sqrt{s^{\,a_1}}\, e^{rs}, \hs \hs w_n =\kappa_2(r) \,\fr{e^{-rs}}{\sqrt{s^{\,a_n}}},\label{enn}
\ee
where $\kappa_i(r)$ for $i=1,2$ are the constants of integration.
The coupled equations for  $u_{k+1}$ and $w_k$ are equivalent (by
expressing $u_{k+1}$ in terms of $w_k$) to the single second-order equation
\be
s^2 \ddot{w}_k +2s \dot{w}_k - \Big (r^2s^2 +(n-1-2k)rs +{n^2 -1 \over 4} \Big )w_k =0, \label{final}
\ee
which may be solved by substituting $w_k=W_k/s$ and $z = 2rs$.
The latter equation is then reduced to the familiar Whittaker equation:
\be
W_{k}^{''} + \Big (- {1\over 4} + {{(2k-n+1 ) \over 2} \over z} +{{1\over 4}-{n^2\over 4}  \over z^2} \Big )W_k =0,
\label{whit}
\ee
where $W_k'' = {d^2 W_k \over dz^2}$.

Therefore, the solutions of (\ref{whit}) are given in a closed form, in terms of  the Whittaker
functions as
\be
W_k = c_1(r)\, M\left(-{n\over 2}  + {1\over 2} +k, {n\over 2}; 2rs\right) + c_2(r)\,
W\left(-{n\over 2}  + {1\over 2} +k, {n\over 2}; 2rs\right) \label{whit2}
\ee
where $c_i(r)$ for $i=1,2$ are constants (see Appendix, for a brief review on Whittaker functions). In Table $1$ and $2$ 
specific examples of the Whittaker $M\left(-{n\over 2}  + {1\over 2} +k, {n\over 2}; 2rs\right)$ and  $W\left(-{n\over 2}  
+ {1\over 2} +k, {n\over 2}; 2rs\right)$ functions are presented.


\begin{center}
{\small\begin{tabular}{|r||r||r||l|}
\hline
 {\tiny $M\left({2k-n+1\over 2}, {n\over 2}; 2rs\right)$}
 & $n=2$\hs \hs \hs &$n=3$ \hs\hs\hs &\hs \hs \hs$n=4$ \\
\hline \hline
$k=1$ \ \ \ \ \ \ \, & $\fr{2^{3/2}\left(\sinh(rs)-rs\,e^{-rs}\right)}{\sqrt{rs}}$ & $\fr{12 \left(rs\cosh(rs)-
\sinh(rs)\right)}{rs}$
&$\fr{12\sqrt{2}\left[3\sinh(rs)-rse^{-rs}+rs(rs-2)\,e^{rs}\right]}{(rs)^{3/2}}$\acc
$k=2$\hs\hs\hs& &$\fr{6\left[\sinh(rs)-rs(1+rs)\,e^{-rs}\right]}{rs}$
&$\fr{12\sqrt{2}\left[rse^{rs}+rs(rs+2)e^{-rs}-3\sinh(rs)\right]}{(rs)^{3/2}}$ \acc
$k=3$\hs\hs\hs & &&$\fr{4\sqrt{2}\left[3\sinh(rs)-rs(3+3rs+2r^2s^2)\,e^{-rs}\right]}{(rs)^{3/2}}$ \acc
\hline
\end{tabular}\acc}
\end{center}


\begin{center}
{\small\begin{tabular}{|r||r||l|}
\hline
 {\tiny $\!\!\!M\left({2k-n+1\over 2}, {n\over 2}; 2rs\right)\!\!\!$}
 & $n=5$\hs \hs \hs \hs\hs\hs &\hs\hs\hs \hs\hs$n=6$  \\
\hline \hline
$k=1$ \ \ \ \ \ \ \, &
$\!\!\!\fr{20\left[3rse^{-rs}+rs(9-6rs+2r^2s^2)\,e^{rs}-12\sinh(rs)\right]}{(rs)^2}\!\!$
& $\!\!\fr{60\sqrt{2}\left[15\sinh(rs)-3rse^{-rs}+rs(r^3s^3-4r^2s^2+9rs-12)\,e^{rs}\right]}{(rs)^{5/2}}\!\!\!\!\!\!$\acc
$k=2$\hs\hs\hs&$\fr{60\left[6\sinh(rs)-rs(rs+3)\,e^{-rs}+rs(rs-3)\,e^{rs}\right]}{(rs)^2}$ \ \ & $\fr{60\sqrt{2}
\left[3rs(rs+4)\,e^{-rs}+rs(2r^2s^2-9rs+18)\,e^{rs}-30\sinh(rs)\right]}{(rs)^{5/2}}$\acc
$k=3$\hs\hs\hs &$\!\!\fr{20\left[3rse^{rs}+rs(2r^2s^2+6rs+9)\,e^{-rs}-12\sinh(rs)\right]}{(rs)^2}$ & $
\fr{60\sqrt{2}\left[30\sinh(rs)+3rs(rs-4)\,e^{rs}-rs(18+9rs+2r^2s^2)\,e^{-rs}\right]}{(rs)^{5/2}}$\acc
$k=4$\hs\hs\hs&
$\!\!\!\!\fr{20\left[6\sinh(rs)-rs(3
+3rs+2r^2s^2+r^3s^3)\,e^{-rs}\right]}{(rs)^2}$ \hs&
$\fr{60\sqrt{2}\left[ 3rse^{rs}+rs(12+9rs+4r^2s^2+r^3s^3)\,e^{-rs}-15\sinh(rs)\right]}{(rs)^{5/2}}$\acc
$k=5$\hs\hs\hs&  &$\fr{12\sqrt{2}\left[15\sinh(rs)-rs\left(15(rs+1)+5r^2s^2(rs+2)+2r^4s^4\right)\,e^{-rs}\right]}
{(rs)^{5/2}}$\acc
\hline
\end{tabular}\acc}
{\bf Table 1:} Explicit expressions for the  Whittaker function $M\left({2k-n+1\over 2}, {n\over 2}; 2rs\right)$
for $n=2,\dots,6$\acc
\end{center}


\begin{center}
\begin{tabular}{|r||r||r||l|}
\hline
 {\tiny $W\left({2k-n+1\over 2}, {n\over 2}; 2rs\right)$}
 & $n=2$\hs \hs &$n=3$ \ \hs\hs &\hs \hs \hs$n=4$ \acc
\hline \hline
$k=1$ \ \ \ \ \ \ \, & $\fr{\left(1+2rs\right)\,e^{-rs}}{\sqrt{2rs}}$ & $\fr{\left(1+rs\right)\,e^{-rs}}{rs}$ \hs\hs
&\hs\hs $\fr{\left(3+2rs\right)\,e^{-rs}}{(2rs)^{3/2}}$\acc
$k=2$\hs\hs\hs& &$\fr{\left[1+2rs(1+rs)\right]\,e^{-rs}}{rs}$
& \hs$\fr{\left[3+2rs(rs+2)\right]\,e^{-rs}}{\sqrt{2}(rs)^{3/2}}$ \acc
$k=3$\hs\hs\hs & &&$\fr{\left[3(1+2rs)+2r^2s^2(3+2rs)\right]\,e^{-rs}}{\sqrt{2}(rs)^{3/2}}$ \acc
\hline
\end{tabular}\acc
\end{center}


\begin{center}
{\small\begin{tabular}{|r||r||l|}
\hline
 {\tiny $W\left({2k-n+1\over 2}, {n\over 2}; 2rs\right)$}
 & $n=5$\hs\hs\hs\hs\hs &\hs\hs\hs\hs \hs$n=6$  \\
\hline \hline
$k=1$ \ \ \ \ \ \ \, &
$\fr{\left(2+rs\right)\,e^{-rs}}{2(rs)^2}$ \hs\hs\hs\hs&
\hs\hs\hs\hs$\fr{\left(5+2rs\right)\,e^{-rs}}{(2rs)^{5/2}}$\acc
$k=2$\hs\hs\hs& $\fr{\left(3+3rs+r^2s^2\right)\,e^{-rs}}{(rs)^2}$\hs\hs \hs&
\hs\hs\hs $\fr{\left(5+4rs+r^2s^2\right)\,e^{-rs}}{\sqrt{2}(rs)^{5/2}}$\acc
$k=3$\hs\hs\hs & $\fr{\left(6+9rs+6r^2s^2+2r^3s^3\right)\,e^{-rs}}{(rs)^2}$ \hs\hs&
\hs\hs $\fr{\left(15+18rs+9r^2s^2+2r^3s^3\right)e^{-rs}}{\sqrt{2}(rs)^{5/2}}$\acc
$k=4$\hs\hs\hs&$\fr{2\left(3+6rs+6r^2s^2+4r^3s^3+2r^4s^4\right)\,e^{-rs}}{(rs)^2}$ &
\hs $\fr{\sqrt{2}\left(15+24rs+18r^2s^2+8r^3s^3+2r^4s^4\right)e^{-rs}}{(rs)^{5/2}}$ \acc
$k=5$\hs\hs\hs&  &
$\fr{\sqrt{2}\left(15+30rs(1+rs)+20r^3s^3+10r^4s^4+4r^5s^5\right)e^{-rs}}{(rs)^{5/2}}$\acc
\hline
\end{tabular}\acc}
{\bf Table 2:} Explicit expressions for the  Whittaker function $W\left({2k-n+1\over 2}, {n\over 2}; 2rs\right)$
for $n=2,\dots,6$\acc
\end{center}

The next step is to derive an orthonormal basis of solutions, and then calculate the Higgs field (\ref{Higgs}).
An orthogonal basis of solutions is given by
\be
\vv_1 =\left( \begin{array}{c}
      v_1\\
      0 \\
      0\\
	  \vdots \\
0\\	
0
\end{array} \right)\,,\hs \vv_{k+1} =\left( \begin{array}{c}
      0\\
      \vdots \\
	  w_k \\
	  u_{k+1}\\
\vdots\\
0
\end{array} \right) \,, \hs\vv_{n+1} =\left( \begin{array}{c}
      0\\
      0 \\
	  \vdots \\
	  0\\
0\\
w_n
\end{array} \right) \,,\hs k\in \{1, \ldots, n-1\}. \label{ortho}
\ee
Introduce the inner product
\bea
<\vv_\kappa,\vv_\lm> &=&\int_{0}^{n+1} \vv_\kappa^\dg\, \vv_\lm\ ds\nonumber\acc
&=&{\cal N}_\kappa\, \delta_{\kappa\lm}  \label{normal}
\eea
then the required solutions of (\ref{Weyl}) are those which are normalizable with respect to (\ref{wnor}).
It is clear that the space of normalizable solutions of (\ref{Weyl}) has dimension $n+1$.

If $\hat{\vv}_1,\dots,\hat{\vv}_{n+1}$ consist an orthonormal basis for  this space then the Higgs field is an  $SU(n+1)$ 
diagonal matrix whose elements are given by
\bea
\Phi_{\kappa} =-i \int_{0}^{n+1} (s-n)\, \hat{\vv}_\kappa^\dg \,\hat{\vv}_\kappa \ ds, \hs \kappa\in \{1, \ldots, n+1\}.
\eea
Recall that, the orthogonal vectors $\vv_1$ and $\vv_{n+1}$ are given in terms of the functions $u_1$ and $w_n$ in  (\ref{enn}). 
The vectors $\vv_{k+1}$ for $k\in \{1,\dots, n-1\}$ are more complicated and are given in terms of the Whittaker functions 
$M\left({2k-n+1\over 2}, {n\over 2}; 2rs\right)$ and $W\left({2k-n+1\over 2}, {n\over 2}; 2rs\right)$. In particular, 
the $w_k$ and $u_{k+1}$ functions are given explicitly  by
\be
w_k = {1 \over s} \, W_k, \hs \hs u_{k+1} = {1\over C_k} \left[\dot{W_k} + \left({a_k-2\over 2s} +r\right)W_k\right] \label{eq2}
\ee
where $W_k$ is defined in (\ref{whit2}).

Performing the required integrals (\ref{normal}), and using (\ref{enn}), (\ref{eq2}), gives
\bea
&& {\cal N}_1 = \int_{0}^{n+1} \sqrt{s^{a_1}}\, e^{rs}\ ds, \nonumber\\
&& {\cal N}_{k+1}= {1\over C_k^2}\left [ \fr{1}{2} \fr{d W_k^2}{ds} +\left(r+\fr{n-1-2k}{2s}\right)\,W_k^2  
\right] {\Big |}_{s=0}^{n+1},\hs \ k \in \{1, \ldots, n-1 \},
\nonumber\acc
&& {\cal N}_{n+1} =\int_{0}^{n+1}  \fr{e^{-rs}}{\sqrt{s^{a_n}}}\ ds.  \label{norm2}
\eea
Note that the normalization factors ${\cal N}_k$, solely consist of boundary terms and thus can be easily evaluated.
However, the Whittaker $W\left({2k-n+1\over 2}, {n\over 2}; 2rs\right)$ function at $s=0$  tends to infinity (see Appendix). 
Hence setting  $c_2(r)=0$ in (\ref{whit2}) to avoid the divergencies, the solutions $W_k$
are given in terms of the Whittaker $M\left({2k-n+1\over 2}, {n\over 2}; 2rs\right)$ function only.

Then the corresponding diagonal elements of the Higgs field at $(0,0,r)$ are given by
\bea
&& \Phi_1 =-{i\over {\cal N}_1} \int_{0}^{n+1} (s-n)\, \sqrt{s^{a_1}}\, e^{rs}\ ds, \nonumber\acc
&&\Phi_{k+1}= -{i \over {\cal N}_{k+1}\,C_k^2}\left [\fr{s}{2}\fr{d W_k^2}{ds}  +\left(rs+{n \over 2}-k-1\right)W_k^2 \right ]
{\Big |}_{s=0}^{n+1}
 \nonumber\\
&& \hs\hs \hs -{i\over {\cal N}_{k+1}\, C_k^2} \int_{0}^{n+1}
\left(\fr{2k-n+1}{2s}-r\right)
W_k^2\,\,ds  + i\,n,  \hs k \in \{1, \ldots, n-1\}, \nonumber\acc
&& \Phi_{n+1} =-{i\over {\cal N}_{n+1}} \int_{0}^{n+1} (s-n)\,\fr{e^{-rs}}{\sqrt{s^{a_n}} }\ ds. \label{higgs2}
\eea
It is a simple task to verify that this solution has indeed the correct asymptotic behavior, and also to recover the 
results obtained for the special cases $n=2$ and $n=3$ in \cite{dancer,HS}, respectively.

To conclude, with the method applied in this section starting from the $n$-dimensional representation of the $SU(2)$ 
algebra ($n$-monopoles)
one ends up with $n+1$ orthonormal vectors, which provide the Higgs field
corresponding to the  minimal symmetry breaking of $SU(n+1)$.
Note that we have focused here on spherically symmetric monopoles located at $(0,0,r)$.
Similarly, the azimuthal dependence can be implemented in a quite straightforward manner via suitable similarity transformations,
or by solving the full Weyl equation with the simple Nahm data. Then the gauge potentials can also be recovered.
 This is a mathematically and physically interesting problem,
and will be addressed in full detail in a forthcoming publication.

\section{Non-Commutative $SU(n+1)$ BPS  Monopoles}

The idea of non-commutative space-time offers a smooth way to introduce nonlocality into field theories without loosing control. 
Motivated by string theory the investigation of non-Abelian gauge theories defined on non-commutative space-time is of great 
interest in the last few years.
In particular, the dynamics of non-Abelian gauge fields involves non-pertrubative field configurations (e.g. instantons, 
monopoles, etc) in an essential way. Thus, in order to quantize a gauge theory it is mandatory to study its classical 
solutions and characterize their moduli spaces.

In \cite{bak} Bak derived a deformed Nahm equation for the BPS equation in the non-commutative $N=4$ super-symmetric
$U(2)$ Yang-Mills theory.
This way, he was able to explicitly construct a monopole solution of the non-commutative BPS
equation to the linear order of the non-commutative scale.

In this section, following Bak, the non-commutative  $SU(n+1)$ BPS spherical symmetric monopoles  with minimal symmetry breaking,
are obtained using a generalization of the ADHMN method.
The non-commutative construction goes along the same lines as the commutative one, with the only difference that
all the ordinary products are now replaced by the associative (but not commutative) Moyal star product ($\star$-product).
The latter is characterized by a constant positive real parameter $\theta$, which appears in the star product of two
functions in the following way
\be
a({\bf x})\star b({\bf x}) = \Big ( e^{{i\over 2}\,\theta^{\mu \nu}\partial_{\mu}\partial'_{\nu}}\, a({\bf x})b({\bf x'})\Big )
{\Big \vert}_{{\bf x}={\bf x'}}.
\ee
For $\theta=0$ it reduces to the ordinary product of functions.
In general,
$\theta^{\mu\nu}$ is a constant, real-value antisymmetric  ($\theta_{\mu\nu} =- \theta_{\nu\mu}$) matrix of order $d$ (where $d$
is the dimension of space-time). For simplicity, we choose its only non-zero terms to be $\theta_{12} = -\theta_{21} = \theta$.

In the non-commutative case, the derivation goes through once all the product operations are replaced by $\star$-product.
Namely, the Weyl equation (\ref{Weyl}) is modified as
\be
\left( -\fr{d}{ds}+\I_n\otimes x_j\sg_j-iT_j\otimes\sg_j\right) \star \vv({\bf x},s)=0. \label{Weyld}
\ee
$T_i$ satisfy (for a more detailed discussion on these issues see \cite{bak}) the {\it deformed} Nahm equation:
\be
{dT_i\over ds} ={1\over 2} \varepsilon_{ijk} [T_j,\ T_k] - \theta \delta_{i3}\,\I.\label{mNahm}
\ee
Setting $T_i = \tilde T_i - \theta s \delta_{i3},\ $ is easy to see that $\tilde T_i$ satisfy the Nahm equations (\ref{Nahm});
they have simple poles; and their residues are irreducible representations of $SU(2)$.

The {\it deformed} Weyl equation takes then the following form
\be
\left( -\fr{d}{ds}+\I_n\otimes x_j\sg_j-i\tilde T_j\otimes\sg_j + \theta \,\hat O\right) \vv({\bf x},s)=0, \label{mWeyl}
\ee
where
\be
\hat O = {i \over 2 }\, \I_n \otimes\left(\sigma_1\partial_2 - \sigma_2 \partial_1\right) + i s \,\I_n \otimes \sigma_3.\ee

Next we construct the actual solutions of the non-commutative BPS equations from the {\it deformed} Nahm data $T_i$ (\ref{mNahm}).
As before, we focus on the spherically symmetric case by setting $x_i =(0,0,r) $, and  $\tilde T_i$ are the ordinary
Nahm data of the form
(\ref{data}) where $f_i = f = -{1\over s}$. In this case, the  {\it deformed} Weyl equation (\ref{mWeyl}) becomes
\be
\left (-{d\over ds} -{f\over 2} \tau_i \otimes \sigma_i + \tilde r \,\I \otimes \sigma_3 \right ) \vv(\tilde r, s)= 0,
\label{dweyl2}
\ee
where $\tilde r = r +i\theta s$.

Similarly to the commutative case,  the {\it deformed} Weyl equation (\ref{dweyl2}) leads to a first-order system of
differential equations  given by  (\ref{equations}) for $r \to \tilde r $.
Then the first and last equations are decoupled and their solutions are given by
\be
u_1 = k_1(\rt) \sqrt{s^{a_1}}\, e^{\left(rs + {i\theta \over 2}s^2\right)},
\hs \hs w_n =  k_2(\rt) \,\fr{e^{-\left(rs + {i\theta \over 2}s^2\right)}}{\sqrt{s^{a_n}}}.
\ee
Finally, the coupled equations for $u_{k+1}$ and $w_k$ are equivalent to the {\it deformed} single second-oder differential equation
\bea
s^2 \ddot{w}_k +2s \dot w_k - \left(\tilde r^2 s^2 -i \theta s^2 +\left(n-2k-1\right)\tilde r s +{n^2 -1\over 4}\right)w_k
=0,\label{dw}
\eea
for  $k \in \{1,\dots, n-1\}$.
Note that, the aforementioned  equations become the corresponding equations of the previous section when $\theta=0$.

The latter equation is solved in terms of the so-called Heun B functions
denoted henceforth as $H_B$ (see, Appendix,
for more information regarding  $H_B$ and the corresponding differential equation). The solution is given by
\bea
w_k \!\!\!\!&=&\!\!\!\! c_1(r) \sqrt{s^{n-1}}\, e^{\left(rs+\fr{i\theta}{2}s^2\right)} H_B
\left(n,\fr{2(-1)^\fr{1}{4}r}{\sqrt{\theta}},
 n-2k-2, \fr{(-1)^\fr{1}{4}(4k +2-2n)r}{\sqrt{\theta}}; (-1)^\fr{3}{4}\sqrt{\theta}s\right)
\nonumber\\
&+&\!\! \!\!\fr{c_2(r)}{\sqrt{s^{n+1}}}\,
 e^{\left(rs+\fr{i\theta}{2}s^2\right)}
 H_B \left(-n,\fr{2(-1)^\fr{1}{4}r}{\sqrt{\theta}},
n-2k-2, \fr{(-1)^\fr{1}{4}(4k+2-2n)r}{\sqrt{\theta}}; (-1)^\fr{3}{4}\sqrt{\theta}s\right)\nonumber\\
 \label{heun}
\eea
where $(-1)^\fr{1}{4}=\fr{1+i}{\sqrt{2}}$ and  $(-1)^\fr{3}{4}=\fr{-1+i}{\sqrt{2}}$. 
It can be shown (see Appendix) that for the specific form of the solutions given by (\ref{heun}) 
the corresponding Heun B functions satisfy the conditions
\be
H_B \left(-n\right)=\left[(-1)^\fr{3}{4}\sqrt{\theta}s\right]^n H_B(n).\ee
Therefore, the general solution  (\ref{heun}) becomes
\be
w_k = c(r, \theta) \sqrt{s^{n-1}}\, e^{\left(rs+\fr{i\theta}{2}s^2\right)}
H_B
\left(n,\fr{2(-1)^\fr{1}{4}r}{\sqrt{\theta}},
 n-2k-2, \fr{(-1)^\fr{1}{4}(4k +2-2n)r}{\sqrt{\theta}};
(-1)^\fr{3}{4}\sqrt{\theta}s\right)\label{fin}
\ee
where $c(r,\theta)$ is an arbitrary  function of $r$ and
$\theta$.
Table $3$ presents the series expansion of the  Heun B function  $H_B\left(n,\fr{2(-1)^{1/4}r}{\sqrt{\theta}},
 n-2-2k, \fr{(-1)^{1/4}(4k +2-2n)r}{\sqrt{\theta}}; (-1)^{3/4}\sqrt{\theta}s\right)$ up to third order: $O(s^3)$.

\begin{center}
{\small
\begin{tabular}{|r||r||r||l|}
\hline
$\!\!\!H_B(n)\propto O\left(s^3\right)\!$
 & $n=2$\hs \hs \hs\hs &$n=3$ \hs\hs\hs\hs &\hs \hs \hs$n=4$ \\
\hline \hline
$k=1$ \ \ \ \ \ \ \, & $\!\!\!1-\fr{4}{3}rs+\left(r^2-\fr{3}{4}I\theta\right)s^2$ & $1-rs+
\fr{3}{5}\left(r^2-I\theta\right)s^2\!\!$
&$\!\!1-\fr{4}{5}rs+\left(\fr{2}{5}r^2-\fr{1}{2}I\theta\right)s^2$\acc
$k=2$\hs\hs\hs& &$\!\!\!1-\fr{3}{2}rs+\fr{2}{5}\left(3r^2-2I\theta\right)s^2\!\!$
&$1-\fr{6}{5}rs+\left(\fr{4}{5}r^2-\fr{2}{3}I\theta\right)s^2\!\!$ \acc
$k=3$\hs\hs\hs & &&$1-\fr{8}{5}rs+\left(\fr{4}{3}r^2-\fr{5}{6}I\theta\right)s^2\!\!$ \acc
\hline
\end{tabular}\acc}
{\bf Table 3:} The $ H_B \left(n,\fr{2(-1)^{1/4}r}{\sqrt{\theta}},
 n-2-2k, \fr{(-1)^{1/4}(4k +2-2n)r}{\sqrt{\theta}}; (-1)^{3/4}\sqrt{\theta}s\right)$ function up to order
 $O(s^3)$ for $n=2,\dots,4$.\acc
\end{center}

Next in order to obtain the {\it deformed} Higgs field we follow the procedure of the previous section.
First we  choose the {\it deformed} orthogonal basis similarly to (\ref{ortho}).
Then the normalization factors are equal to
\bea
&& {\cal N}_1 = \int_{0}^{n+1} \sqrt{s^{a_1}}\, e^{\left(rs+ {i\theta\over 2}  s^2\right)}\ ds, \nonumber\\
&& {\cal N}_{k+1}= {1\over C_k^2}\left [\fr{1}{2} \fr{d W_k^2}{ds}  +\left(r +i \theta s+\fr{n-1-2k}{2s}\right)W_k^2
\right ]{\Big |}_{s=0}^{n+1}, \hs k \in \{1, \ldots, n-1 \},\nonumber\\
&& {\cal N}_{n+1} =\int_{0}^{n+1} \fr{e^{-\left(rs+{i\theta\over 2}  s^2\right)}}{ \sqrt{s^{a_n}}}\ ds,
\eea
where $W_k =s w_k$ and $w_k$ is given by (\ref{fin}).
The corresponding diagonal elements of the Higgs field are then expressed as
\bea
&& \Phi_1 =-{i\over {\cal N}_1} \int_{0}^{n+1} (s-n)\sqrt{s^{a_1}}\, e^{\left (rs+{i\theta\over 2}  s^2\right)}\ ds, \nonumber\acc
&& \Phi_{k+1}= -{i \over {\cal N}_{k+1}C_k^2}
\left [\fr{s}{2} \fr{d W_k^2}{ds}  +  \left(rs +i\theta s^2+{n \over 2}-k-1\right)W_k^2
\right ]{\Big |}_{s=0}^{n+1}\nonumber\\
&&\hs\hs\hs\! -{i\over {\cal N}_{k+1} C_k^2 } \int_{0}^{n+1}
\left(\fr{2k-n+1 }{2s}-r-i\theta s\right)
 W^2_k\ ds + i\,n,\hs k \in \{1, \dots, n-1 \},\nonumber\acc
&& \Phi_{n+1} = -{i\over {\cal N}_{n+1}} \int_{0}^{n+1} \left(s-n\right)\,
\fr{e^{-\left(rs+{i\theta\over 2}  s^2\right)}}{\sqrt{s^{a_n}} }\ ds.
\eea
As expected, the expressions above reduce to equations (\ref{norm2}) and (\ref{higgs2}) when the deformation parameter vanishes, 
i.e. in the limit $\theta\ra 0$.

\section{Discussion}
We have been able to derive, using a quite elegant and unifying methodology,
explicit expressions for spherically symmetric commutative and non-commutative $SU(n+1)$
BPS monopoles in the minimal symmetry breaking case.
The use of the generic  $n$-dimensional representation of $SU(2)$ allowed the explicit construction of solutions
of the Weyl equation giving rise to sets of simple differential equations. The solutions of the aforementioned
differential equations
were expressed in terms of confluent (Whittaker) or biconfluent (Heun B) hypergeomentric functions for the
commutative and the non-commutative case, respectively.

Although we have restricted our investigation
in the spherically symmetric case
it is important to note that azimuthal dependence may
be implemented in the generic situation by means of
suitable similarity transformations,
however this issue will be discussed in full detail in a forthcoming work.
It is also worth noting that the case $n\to \infty $ merits special investigation, given that the behavior
of the discovered solutions in this situation is rather particular, modifying the form
of the final expressions for the Higgs field. Physical interpretation of the behavior of the Higgs
field is also desirable.
Such results may be also of relevance when one considers, within this context,
infinite dimensional representations of
$SL(2,R)$ (the non-compact case, e.g. non-trivial spin $s=0$ representation of $SL(2,R)$).
The solution of the Weyl equation
in this case involves a particular polynomial basis, given that the associated representation is expressed
in terms of simple differential operators. These are quite intriguing issues, and will be left for future investigations.

\vskip 20pt
\centerline{\bf Acknowledgements}
\vspace{.25cm}
\noindent

TI thanks Paul Sutcliffe and Wojtek Zakrzewski for useful discussions.

\appendix
\section{Whittaker and Heun B functions}

The {\it Whittaker functions $M(k,m;z)$ and  $W(k,m;z)$} solve the {\it Whittaker differential equation} (see e.g. \cite{ww, abst}):
\be
{d^2 W \over dz^2} + \left ( -{1\over 4}  + {k\over z} +{{1\over 4} -m^2 \over z} \right ) W =0
\ee
where
\be
W = c_1\, M\left(k,m;z\right)+c_2\,W\left(k,m;z\right).
\ee
They can be defined in terms  of confluent hypergeometric functions as follows
\bea
M\left(k,m,z\right) &=& e^{-{z\over 2}}\,z^{m+{1\over 2}}\  _1F_1 \left({1\over 2}+m-k; 1+2m; z\right), \\
W\left(k,m,z\right) &=& e^{-{z\over 2}}\,z^k\  _2F_0 \left({1\over 2}+m-k, {1\over 2} -m-k; -z^{-1}\right),
\eea
where $_1F_1\left(a;b;z\right)=e^z\ _1F_1\left(b-a; b;-z\right)$ and $\ _n F_m$ are generalized
hypergeometric functions (see e.g. \cite{ww, abst, arfken}) defined as:
\be
_p F_{q}\left(a_1, a_2, \ldots, a_p; b_1, b_2, \ldots b_q; z\right) = \sum_{l=0}^{\infty} {(a_1)_l\, (a_2)_l \ldots (a_p)_l \over
(b_1)_l\, (b_2)_l \ldots (b_q)_l} {z^l \over l!}, ~~~~(a)_l = {\Gamma(a+l) \over \Gamma(a)}.
\ee

In the special case where $m + {1\over 2} -k$ is a positive integer (which is our case), the  Whittaker $W\left(k,m,z\right)$ 
function is defined by the integral function
\be
W\left(k,m;z\right) = {e^{-{z\over 2}} z^k \over \Gamma({1\over 2} - k +m)} \int_{0}^{\infty} t^{-k -{1\over 2} +m}
(1+{t\over z})^{k-{1\over 2} +m} e^{-t}\ dt
\ee
for all values of $z$ except negative real values.
Then, it is obvious that at $z=0$ (corresponding to $s=0$ in our case) it diverges.

Let us now introduce the {\it Heun Biconfluent differential equation}
\be
{d^2 y \over d z^2} + \left (-2z -b +{1+a \over z} \right ) {d y \over dz} + \left
( c-a -2 -{1\over 2}{(a+1)b +d \over z}\right )y =0.\label{hb}
\ee
The Heun B function  $H_B(a,b,c,d;z)$ is a Frobenius solution to Heun's Biconfluent equation  (\ref{hb})
computed as a power series expansion around the origin, which is a regular singular point.
In this case, the growth of solutions is bounded, in any small sector, by an algebraic function.
Because the next singularity is located at infinity, this series converges in the whole complex plane.

Note that the Heun B function can be expressed in terms of the Whittaker $M$ function for specific values of
the parameters, i.e.
\be
H_B(a,0,c,0;z)=\fr{1}{z^{\fr{a}{2}+1}}\,e^\fr{z^2}{2}\,M\left(\fr{c}{4},\fr{a}{4};z^2\right).
\ee
It is thus clear that the results of section $3$ reduce to those of section $2$, in the limit $\theta \ra 0$.

A special case occurs when in $H_B(a,b,c,d,z)$, the third parameter satisfies the condition $c=2(k+1)+a$   where $k$ 
is a positive integer (which is our case).
 In this case the $k^{\mbox{th}}+1$ coefficient in the series expansion is a polynomial of degree $k$ in $d$. When $d$ 
 is a root of this polynomial, the $k^{\mbox{th}}+1$ and subsequent coefficients cancel and the series truncates resulting 
 in a polynomial form of degree $k$ for Heun B.


\end{document}